\documentclass[prb,superscriptaddress,showpacs,twocolumn]{revtex4-1}

\usepackage{graphicx}
\usepackage{color}
\usepackage{graphicx}
\usepackage{amssymb}
\usepackage{epstopdf}
\usepackage{times}
\usepackage{ulem}
\def\6r3{$(6\sqrt{3}\times 6\sqrt{3})$R30$^\circ$}

\begin{document}


\title{Electronic structure of graphene on single crystal copper substrates} 
\date{\today}


\author{Andrew L. Walter$^{1,2,a}$, Shu Nie$^{3}$, Aaron Bostwick$^{1}$, Keun Su Kim$^{1,4}$, Luca Moreschini$^{1}$, Young Jun Chang$^{1,2}$, Davide Innocenti$^{5}$, Karsten Horn$^{2}$, Kevin F. McCarty$^{3}$, Eli Rotenberg}

\address{ Advanced Light Source (ALS), E. O. Lawrence Berkeley National Laboratory, Berkeley, California 94720, USA,\\ $^{2}$ Department of Molecular Physics, Fritz-Haber-Institut der Max-Planck-Gesellschaft, Faradayweg 4-6, 14195 Berlin, Germany,\\ $^{3}$ Sandia National Laboratories, Livermore, California 94550, USA.\\ $^{4}$Center for Atomic Wires and Layers, Pohang University of Science and Technology, Pohang 790-784, Korea.\\ $^{5}$University of Rome (Tor Vergata), Rome 00173, Italy \\ $^{a}$ email: alwalter@lbl.gov}



\date{\today}

\def\EF{$E_{\mathrm{F}}$}
\def\ED{$E_{\mathrm{D}}$}
\def\kk{K-K}
\def\gk{$\Gamma$-K}
\def\gkm{$\Gamma$-K-M}
\definecolor{Andrew}{rgb}{1,0,0}
\def\Andrew{\textcolor{Andrew}}
\definecolor{eli}{rgb}{0,.5,1}
\def\eli{\textcolor{eli}}
\def\skw{$\Sigma(E, {\bf k})$}
\def\Ga{$ \mathrm{G}_{111}$}
\def\ga{$ \mathrm{G}_{100}^ \mathrm{A}$}
\def\gb{$ \mathrm{G}_{100}^ \mathrm{B}$}
\def\EGa{$E_{G_{111}}$}
\def\Ega{$E_{G_{100}^A}$}
\def\Egb{$E_{G_{100}^B}$}

\begin{abstract}
The electronic structure of graphene on Cu(111) and Cu(100) single crystals is investigated using low energy electron microscopy, low energy electron diffraction and angle resolved photoemission spectroscopy. On both substrates the graphene is rotationally disordered and interactions between the graphene and substrate lead to a shift in the Dirac crossing of $\sim$ -0.3 eV and the opening of a $\sim$ 250 meV gap. Exposure of the samples to air resulted in intercalation of oxygen under the graphene on Cu(100), which formed a ($\sqrt{2} \times 2\sqrt{2}$)R45$^{\rm o}$ superstructure. The effect of this intercalation on the graphene $\pi$ bands is to increase the offset of the Dirac crossing ($\sim$ -0.6 eV) and enlarge the gap ($\sim$ 350 meV). No such effect is observed for the graphene on Cu(111) sample, with the surface state at $\Gamma$ not showing the gap associated with a surface superstructure. The graphene film is found to protect the surface state from air exposure, with no change in the effective mass observed, as for 1 monolayer of Ag on Cu(111).

\end{abstract}

\pacs{}

\maketitle 


\section{Introduction}
Graphene's unique electronic properties\cite{Geim:2007p4020} have excited intense research interest since its isolation by micromechanical cleavage in 2004\cite{Novoselov:2004p5403}. However this technique, used in these early investigations, is unsuitable for large-scale production of graphene sheets. Thermal decompositon of SiC\cite{Deheer:2007p5404,Rutter:2007p5405,Forbeaux:1998ic,Emtsev:2008p4975}, surface segregation of C dissolved in metal substrates\cite{Sutter:2008p5406,Yu:2008p5407,McCarty:2009p5408} and decomposition of hydrocarbons at metal surfaces\cite{Loginova:2009p5409,Ndiaye:2006p5410} have all been shown to produce graphene and are potentially scalable to commercial production. 

Recently the growth of large scale graphene films ($\sim $ 1 m wide) has been accomplished using chemical vapor deposition (CVD) onto polycrystalline Cu foils\cite{Bae:2010es}, leading to a strong interest in the electronic and structural properties of the graphene/Cu interface.  The structural properties of graphene/Cu was investigated\cite{Wofford:2010p5394} in detail, however electronic structure data, particularly Angle Resolved Photoemission Spectroscopy (ARPES), on such films are hampered by the many randomly oriented domains in both the graphene and the substrate. This leads to a large collection of electronic bands from both the substrate and the graphene, greatly confusing the interpretation of the ARPES data.

A recent attempt to overcome this problem by Varykhalov et al. involved measuring graphene on 1 monolayer of Cu/Ni(111), which indicated that the interaction between graphene and the substrate was significantly underestimated by current DFT L(S)DA theory\cite{Varykhalov:2010fd}. The method involved growing a graphene film on a 15-20 monolayer Ni(111) film on W(110) followed by intercalating a monolayer of Cu(111) underneath the graphene. While providing insight into the electronic structure of graphene on Cu, it is an open question whether this arrangement reflects the properties of the interface between graphene and bulk copper. The growth of graphene on single crystal Cu, undertaken in the current study, provides an ideal system for investigating the electronic structure of graphene on Cu by eliminating the rotational disorder in the substrate. As the bulk and surface Cu bands are well known, in our samples it is easy to identify the graphene $\pi$ bands. 

Graphene films were grown on Cu(111) and Cu(100) single crystals and characterized using low energy electron microscopy (LEEM) and low energy electron diffraction (LEED). The electronic structure of the graphene samples was investigated using ARPES measurements. Here we show that, similar to  graphene on 1ML Cu/Ni(111)\cite{Varykhalov:2010fd}, graphene on Cu (100) and (111) single crystals is $n$-doped with the Dirac crossing energy \ED\ located -0.3 eV below the Fermi level \EF. We find a gap of $\sim 250$ meV at the Dirac crossing of the graphene $\pi$ bands, significantly larger than the gap in graphene  on 1ML Cu/Ni(111)  ($\sim$ 180 meV) and  the LDA estimate  ($\sim$ 11 meV) \cite{Giovannetti:2007jc}.

We also show that air exposure leads to oxygen intercalation under the graphene on Cu(100), forming a ($\sqrt{2} \times 2\sqrt{2}$)R45$^{\rm o}$ superstructureand giving a larger doping ($\sim$ -0.6 eV Dirac crossing shift) and an increased gap ($\sim$ 350 meV in the graphene $\pi$ bands at \ED). The Cu(111) surface state at $\Gamma$ (Brillouin zone centre) does not show the gap expected for such an intercalted surface structure, confirming that exposure to air does not produce the same effect for graphene on Cu(111). In contrast the graphene protects the surface state at $\Gamma$ from air exposure, with no change in the effective mass observed.

\section{Methods}
 
Cu(100) and Cu(111) single crystals were initially cleaned by annealing at $900^\circ$C for about 12 hours in a flowing mixture of hydrogen and argon at atmospheric pressure followed by sputtering with argon ions. Subsequent annealing at $750^\circ$C in the LEEM produced surfaces with atomically flat terraces separated by monatomic Cu steps and step bunches. Large bunches of steps formed on Cu(111) at pinning sites spaced about 2 - 5 microns apart. Carbon was deposited onto the substrates at $850 - 900^\circ$C from a graphite rod heated by an electron beam. The growth was imaged in real time using LEEM. The deposition was stopped near the completion of the first graphene layer using the ability of LEEM to distinguish bare Cu, monolayer graphene and bilayer graphene\cite{Ohta:2008p262,Hibino:2009gh}.
 
Selected-area LEED patterns were acquired after growth from regions 2 or 20 $\mu$m in diameter. The graphene-covered Cu crystals were removed from the ultrahigh vacuum of the LEEM system and exposed to air for about 3 hours before again being placed in vacuum and degassed at about $350^\circ$C for ARPES measurements.  After the ARPES measurements, the samples were re-introduced into the LEEM for further analysis.  Both the graphene/Cu(100) and graphene on Cu(111) specimens experienced an additional 3 hours exposure to air while the graphene/Cu(111) was also kept in a desiccator for several weeks prior to analysis. The samples were briefly degassed at about $250^\circ$ C before the subsequent LEEM/LEED measurements. The LEEM/ LEED measurements where performed at Sandia National Laboratories in Livermore, California.
 
ARPES spectra were obtained at the Electronic Structure Factory endstation (SES-R4000 analyzer) at beamline 7 of the Advanced Light Source, Lawrence Berkeley National Laboratory. A photon energy of 95 eV was used giving overall resolutions of $\sim$ 25 meV and $\sim0.01 \mathrm{\AA}^{-1}$. The spatial area sampled by the  ARPES spectra was typically 50 to 100 $\mu$m. During the measurements the samples were cooled to $\sim$ 20 K using a liquid He-cooled cryostat and the pressure was $<$  2 $\times$10$^{-10}$ Torr.

The ARPES spectrum of graphene on Cu single crystals was modelled as follows. The bare $\pi$ band dispersion \ $E_\mathrm{bare}({\bf k})$ was computed using a first nearest neighbour tight binding (TB) model based on the approach given by Saito et al.\cite{Saito:1998p5382}. The secular equation $|H-E_\mathrm{bare\pm}({\bf k}).S|=0$ is solved for the eigenvalues $E_\mathrm{bare\pm}({\bf k})$, with

\[ H
= \left(
\begin{array}{cc}
\epsilon_{2p} - E_{gap}/2 & \gamma_{0} f(k) \\ 
\gamma_{0}f(k)^{*} & \epsilon_{2p} - E_{gap}/2 \\
 \end{array}
\right)\] \begin{equation} \end{equation}
\[ S
= \left(
\begin{array}{cc}
1 & s_{0} f(k) \\ 
s_{0}f(k)^{*} & 1 \\
 \end{array}\right)\]

and
\begin{equation}
f(k) = \exp(ik.R_{1}+ik.R_{2}+ik.R_{3})
\end{equation}
where $\gamma_{0}$ is the hopping potential, $s_{0}$ is the overlap potential and $R_{1}, R_{2}, R_{3}$ are the vectors denoting the position of the 3 first nearest neighbor carbon atoms.

A lattice constant, $a = 2.46$, is employed as are the fitted parameters, $\gamma_{0}$=-3.24 eV and $s_0$=0.0425 eV, determined  by Bostwick et al. for graphene on \6r3 C- SiC(0001) \cite{Bostwick:2007p252}. The fitted parameter, $\epsilon_{2p}$, is the offset of the Dirac energy, \ED\ , from the Fermi level due to doping of the graphene by the substrate and is determined by comparison to the graphene on Cu ARPES measurements. The $E_{gap}$ parameter is introduced by us to account for the introduction of an energy gap at \ED\ due to the breaking of graphene's A-B lattice symmetry.  Broadening of the bare band is introduced to the model through the self-energy via the single-particle spectral function:

\begin{equation}
 A(E, {\bf k}) = \frac{\scriptstyle |{\rm Im} \Sigma(E, \bf k)|}{\scriptstyle (E-E_{bare}({{\bf k}})-{\rm Re}\Sigma(E,{\bf k})^2)+{\rm Im}\Sigma(E,{\bf k})^2}.
\end{equation}
 
\begin{figure}
\includegraphics{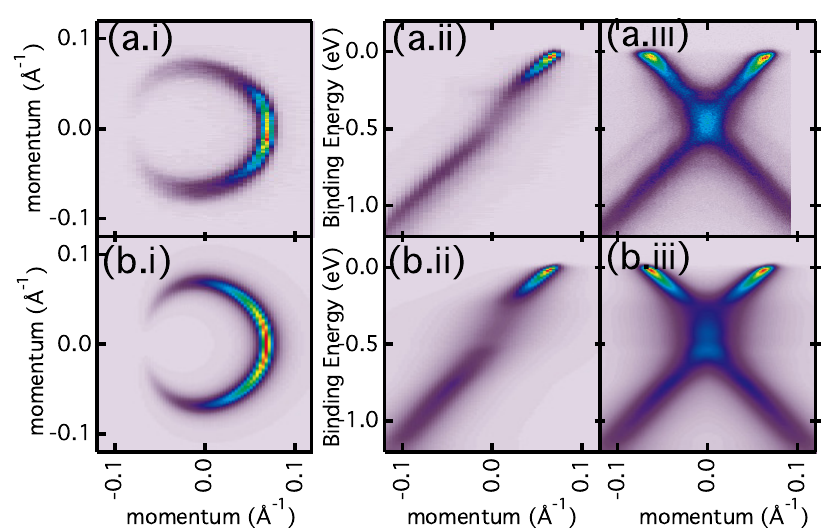}
\caption{\label{model}Experimental fermi surface and $\pi$ band structure obtained from graphene on \6r3 C-SiC (a i, ii and iii) for comparison to a semi empirical model (b i, ii and iii). The model employs a first nearest neighbour tight binding model for the bare (un-broadened) bands with the parameters ($\gamma_{0}$=-3.24 eV, s$_0$=0.0425 eV and $\epsilon_{2p}$=-0.47 eV) determined from a fit to the experimental data. Broadening is performed using the self-energy, obtained by a fit to the experimental data in the $\Gamma$ - K direction, and experimental broadening of 25 meV and 0.01 \AA$^{-1}$. The simulation and experiment agree qualitatively.}	
\end{figure}

The self-energy, \skw\, is determined using the semi-empirical method of Bostwick et al. \cite{Bostwick:2007p247}, where the linewidth of the ARPES data is used to determine the imaginary component of the self-energy, which is Hilbert transformed to get the real component. This experimental self-energy is then used to recreate the experimental data and a self-consistent fitting used to further refine the self-energy. An additional gaussian broadening term is added to the current model to account for experimental broadening in energy, $\Delta E = 25$ meV, and momentum, $\Delta k = 10$ $m$\AA$^{-1}$. Fig. \ref{model} shows experimental data for graphene on \6r3 C-SiC(0001) (data adapted from Bostwick et al.\cite{Bostwick:2007p247}) compared to a model spectral function fitted according to the above procedure. The fitted values, excepting $\epsilon_{2p}$ and $E_{gap}$, used for the graphene on Cu data model are those obtained from this fit as the rotational disorder prohibits fitting in this case.

For comparison to the graphene on single crystal Cu experimental data many rotationally disordered domains need to be modeled. This is done by producing a 3D (2 momentum and 1 energy) spectral function for a single graphene domain, as shown in fig. \ref{model}. This single domain is then rotated around the $\Gamma$ point (in this case 20 single domains, with evenly spaced rotation angles between $\pm$  2.5 $^\circ$) to produce the spectral function from 20 rotationally disordered graphene domains. The rotational disorder model is then produced by summing the spectral function from each of these rotational domains together. Inclusion of additional domains, or domains with larger rotations, did not appreciably change the resulting spectral function, hence these values where employed.

\begin{figure}
\includegraphics{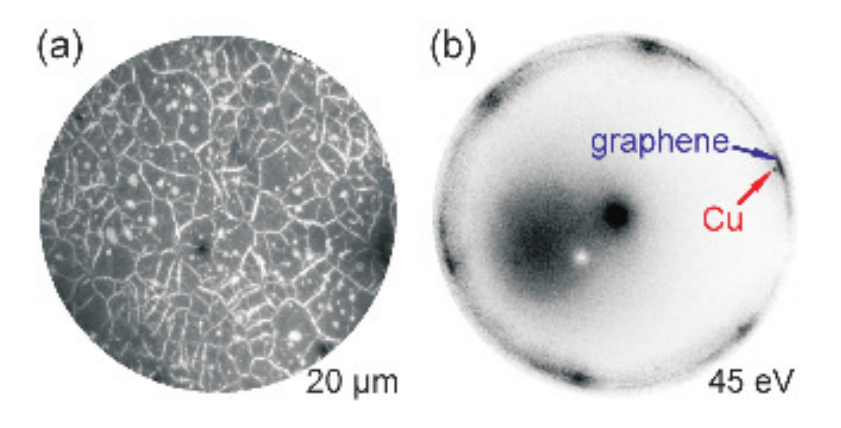}
\caption{\label{Cu111LEEM}LEEM/LEED characterization of a nearly complete single-layer graphene film on Cu(111). (a) LEEM image with 20 $\mu$m field of view. Bright dots are nucleation sites and bright lines are boundaries between graphene islands. (b) LEED from a 20-$\mu$m diameter area of (a). First-order diffraction spot of Cu marked by red arrow and portion of graphene arc marked by blue arrow.}
\end{figure}

\begin{figure*}
\includegraphics{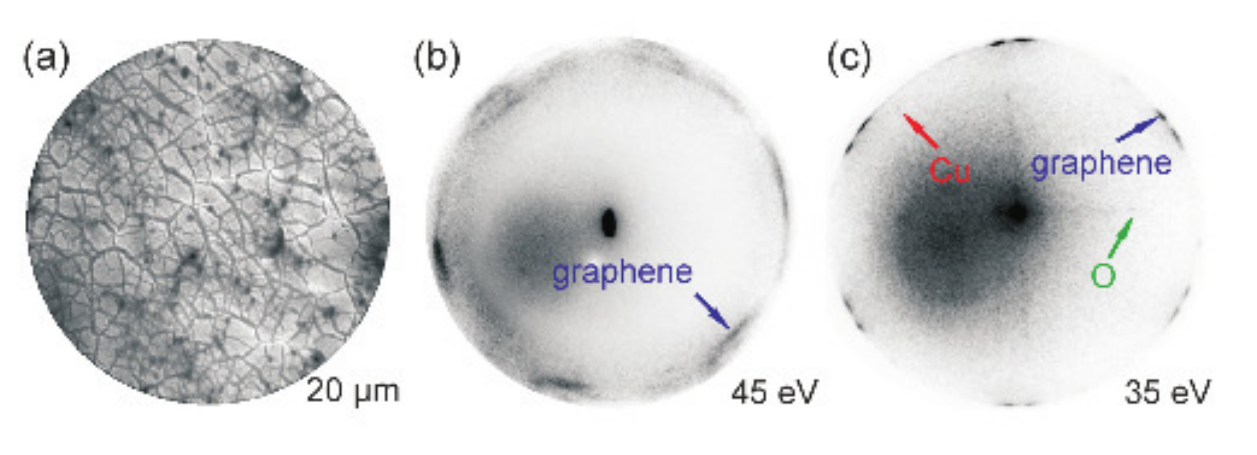}
\caption{\label{Cu100LEEM}LEEM/LEED characterization of single-layer graphene film on Cu(100). (a) LEEM image with 20 $\mu$m field of view. Dark lines are boundaries between graphene islands. (b) LEED from 20 $\mu$m diameter area of as-grown graphene. One graphene arc marked by blue arrow. (c) LEED from a 2 $\mu$m diameter area of air-exposed sample. Diffraction spots of Cu, graphene and intercalated O marked by red, blue and green arrows, respectively.}
\end{figure*}

\begin{figure*}
\includegraphics{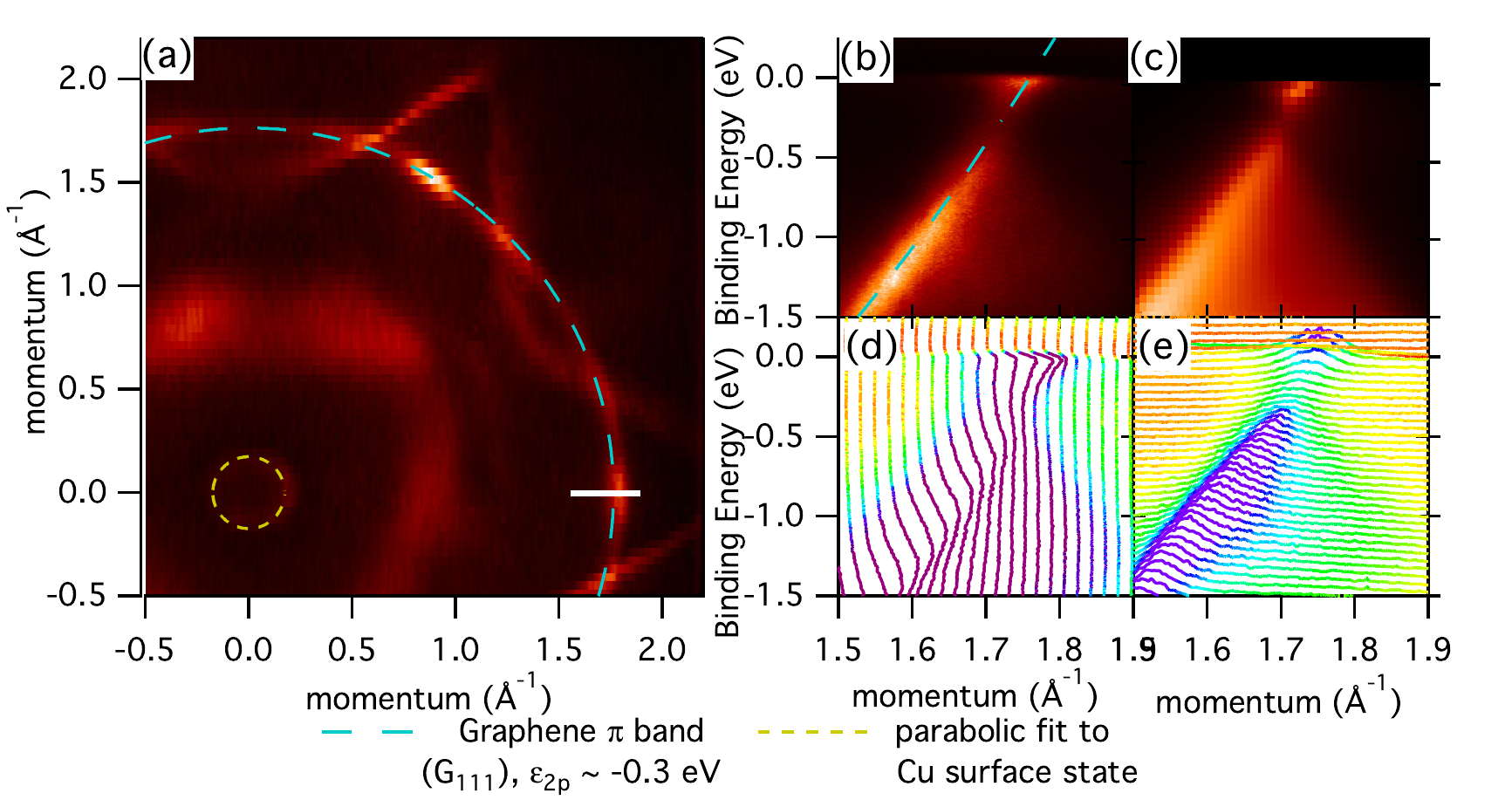}
\caption{\label{Cu111} Experimental Fermi surface (a) obtained from graphene on single crystal Cu(111). Also presented is the experimental spectral function (b), semi emperical model spectral function (c), energy distribution curves (d) and momentum distribution curves (e) obtained along the white line in (a) (Cu Brillouin zone $\Gamma$-K direction). The graphene $\pi$ bands calculated from a first nearest neighbour tight binding model ($\gamma_{0}$=-3.24 eV, s$_0$=0.0425 eV and $\epsilon_{2p}$=-0.3 eV) are overlayed as the large dashed blue lines. A parabolic fit to the Cu(111) surface states is also shown (small dashed yellow lines). All other unmarked bands are due to the bulk Cu. The graphene Fermi surface is observed as an uninterrupted ring due to rotationally disordered graphene domains. The preferential alignment (higher intensity) of the domains along the $\Gamma$ - K direction is also observed in LEED.}
\end{figure*}

\begin{figure}
\includegraphics{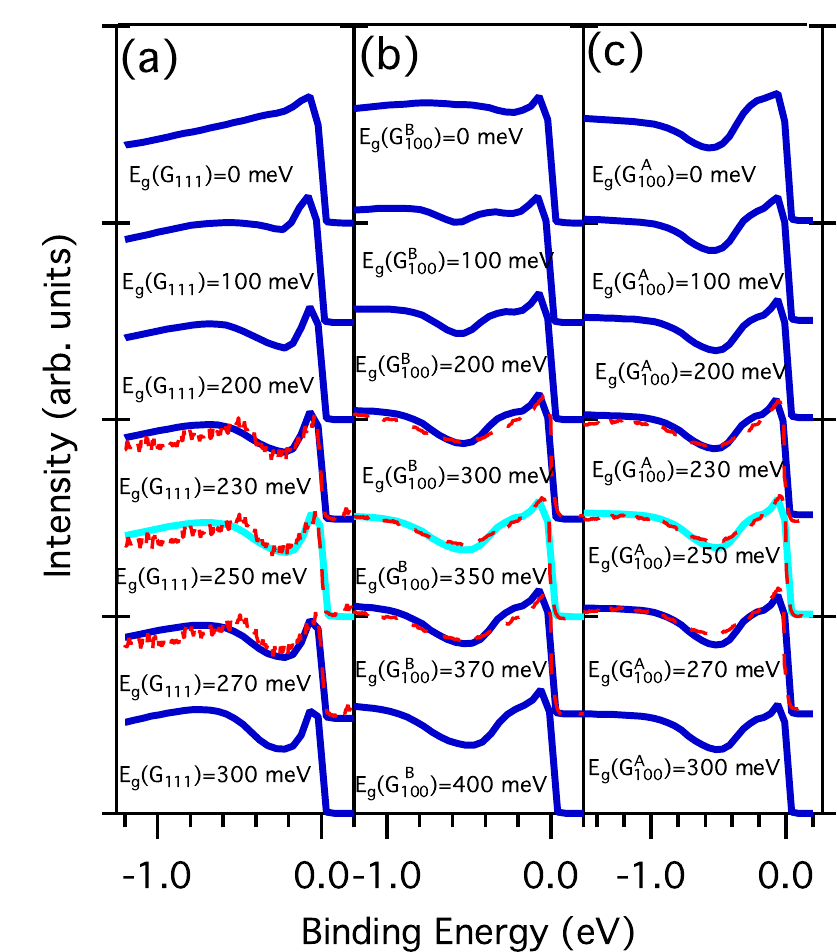}
\caption{\label{Theory}The energy distribution curves (red dashed lines) obtained at the K point from the semi-empirical model for a range of simulated band gaps (blue solid lines) are presented for (a) Cu(111), together with a single band model \Ga\ whose gap $E_g$(\Ga) is varied, and (b, c) for Cu(100), which is compared to a two-band model \ga, \gb.  In (b), the gap $E_g$(\gb) is varied for fixed $E_g$(\ga)= 250 meV, while in (c), the gap $E_g$(\ga) is varied for fixed $E_g$(\gb)=350 meV. In all panels, the light blue lines represent best-fit models to the data.}
\end{figure}

\begin{figure*}
\includegraphics{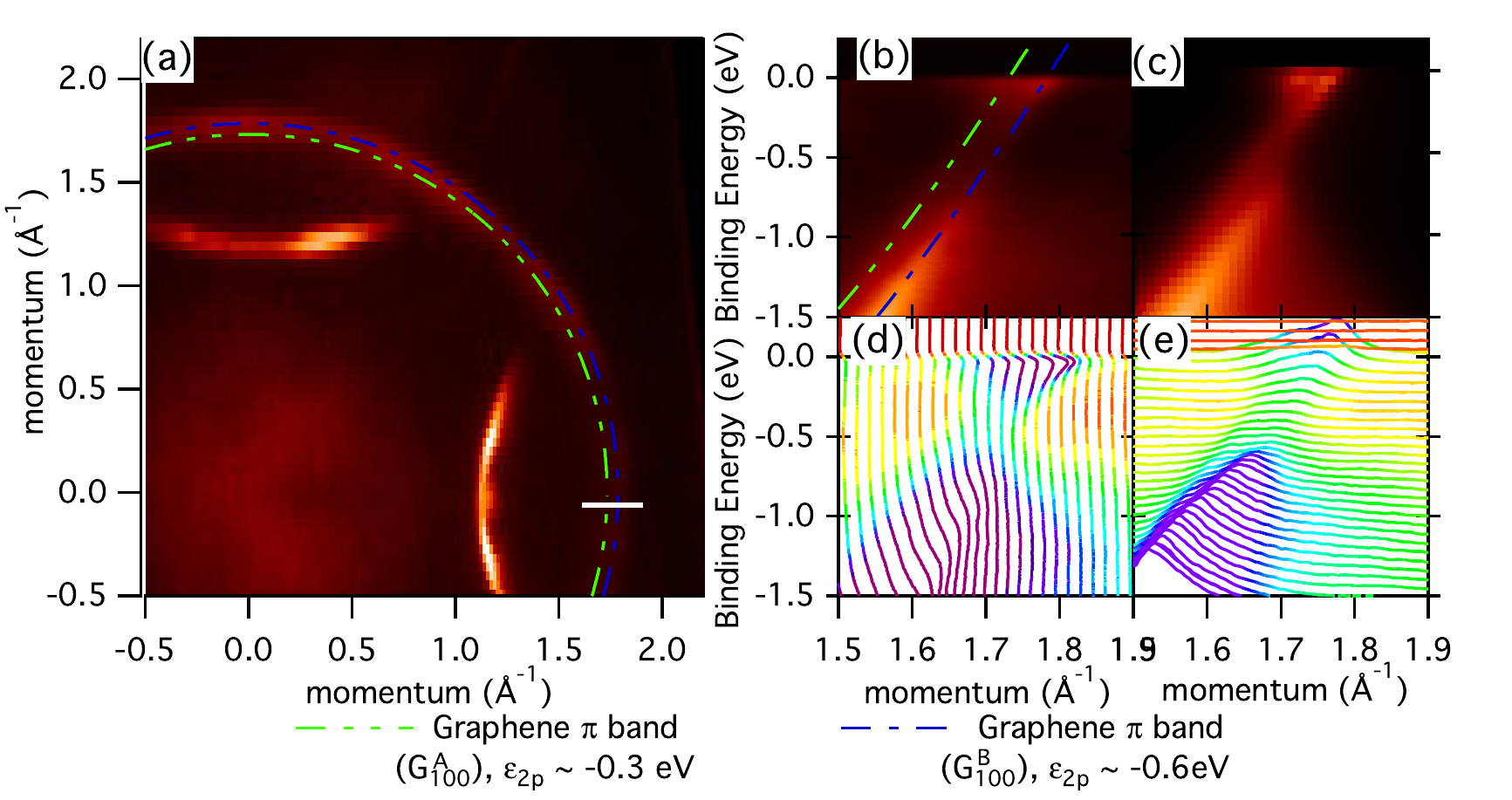}
\caption{\label{Cu100}Experimental Fermi surface (a) obtained from graphene on single crystal Cu(100). Also presented is the experimental spectral function (b), semi emperical model spectral function (c), energy distribution curves (d) and momentum distribution curves (e) obtained along the white line in (a) (Cu Brillouin zone $\Gamma$-K direction). The graphene $\pi$ bands calculated from a first nearest neighbour tight binding model ($\gamma_{0}$=-3.24 eV, s$_0$=0.0425 eV and $\epsilon_{2p}$=-0.3 eV) are overlayed as the dash dot blue and dash dot dot green lines, respectively. All other unmarked bands are due to the bulk Cu. The graphene Fermi surface is observed as an un-interuppted ring due to rotationally disordered graphene domains.}
\end{figure*}

\section{Results and Discussion}
\subsection{LEEM of as-grown films}
Figs. \ref{Cu111LEEM}(a) and \ref{Cu100LEEM}(a) show representative LEEM micrographs of graphene on Cu(111) and Cu(100), respectively. These images show that graphene grows in grains separated by boundaries, whose intensity can be bright or dark depending on the electron energy used for imaging. The bright dots in Fig. \ref{Cu111LEEM} (a) are the graphene nucleation sites. Similar growth structures have been observed in STM measurements of graphene on Cu(100)\cite{Rasool:2011eh}.

Figs. \ref{Cu111LEEM}(b) and \ref{Cu100LEEM}(b) show representative LEED patterns integrated over the 20 $\mu$m regions of Figs. \ref{Cu111LEEM}(a) and \ref{Cu100LEEM}(a) for Cu(111) and (100), respectively. The LEED pattern obtained from graphene on Cu(111) consists not of six sharp spots expected for a perfect graphene sheet, but instead  is smeared into a nearly compete ring. This shows that graphene grows on Cu(111) with a substantial amount of azimuthal disorder. The graphene diffraction is most intense near the first-order Cu(111) spots, indicating a preference for graphene to align with the Cu lattice. Subsequent growth experiments established that the graphene not aligned with the Cu(111) substrate originated from islands that nucleated at the large bunches of Cu steps\cite{Nie:2011tf}. Reducing the bunch density leads to higher fractions of aligned graphene. The graphene LEED pattern is also markedly broadened in the radial direction, a result of the Cu hillocks that form when uncovered Cu sublimes during graphene growth \cite{Wofford:2010p5394}. The broadening of the graphene LEED pattern in the radial direction results from the fact that the graphene conforms to this roughened topography.

The LEED pattern obtained from graphene on the Cu(100) crystal is similar to that from Cu(111) crystals and (100) grains of copper foils. The LEED pattern consists of two sets of broad, six-fold arcs \cite{Wofford:2010p5394}. (One set of arcs is stronger in the example in Fig. \ref{Cu100LEEM}(b)). The two sets of arcs arise because there are two symmetry-equivalent ways to put the six-fold graphene on the four-fold Cu(100) surface. The arcs occur because the graphene has a large range of in-plane orientations. In addition, the arcs are broadened in the radial direction, a result of the roughened surface topology, as for Cu(111) and STM measurements\cite{Rasool:2011eh}. Reducing the integration area for the LEED patterns down to 2 microns does not completely eliminate the azimuthal disorder observed in the LEED patterns for both Cu(100) and Cu(111).  Therefore the size of a domain having a single in-plane orientation is smaller than the typical grains seen in Figs. \ref{Cu111LEEM} (a) and \ref{Cu100LEEM} (a).

\subsection{Angle Resolved Photoemission Spectroscopy}

Angle Resolved Photoemission Spectroscopy (ARPES) measurements of graphene on Cu(111) and Cu(100) surfaces are presented in Figs. \ref{Cu111} and  \ref{Cu100}, respectively.  The Fermi surface of the graphene on Cu(111) is presented in Fig. \ref{Cu111}(a). The large nearly continuous arc (indicated by the large dashed line) corresponds to the Fermi surface of the rotationally disordered graphene $\pi$ bands. Similar to the LEED pattern presented above, an increase in the intensity in the Cu(111) $\Gamma$-K direction is observed. This is the direction along which the two lattices are aligned.The other features in Fig. \ref{Cu111}(a) correspond to cuts through the Cu(111) Fermi surface and the central circular arc of the Cu(111) surface state \cite{Gartland:1975dv,Kevan:1983jz}. Since the clean Cu(111) surface state does not survive exposure to air and since the LEEM images show that the substrate is covered, the surface state observed here must be localized to the interface between graphene and Cu(111).

The experimental spectral function obtained along the white line in Fig. \ref{Cu111}(a) is presented in Fig. \ref{Cu111}(b) with the TB fit obtained using $\epsilon_{2p}$ = -0.3 eV shown in blue (dashed line). The good agreement between the TB fit and the experimental data indicates that the graphene is electron ($n$-type) doped by the Cu layer. The effect of this doping is a shift in the Dirac crossing to -0.3 eV, in good agreement with data obtained from graphene on 1ML Cu/Ni(111)\cite{Varykhalov:2010fd}(-0.31 eV). 

Varykhalov et al.\cite{Varykhalov:2010fd} also found that a gap was opened at the Dirac point ( $\sim$ 180 meV) in the graphene on 1ML Cu/Ni(111) system. At first glance the spectral function in Fig. \ref{Cu111} (b) also suggests a gap for graphene on single crystal Cu however the Energy Distribution Curves (EDC) and Momentum Distribution Curves (MDC) shown in Fig. \ref{Cu111} (d) and (e) indicate intensity is present throughout the Dirac crossing region which suggest the presence of states.  Analysis shows that this intensity is largely caused by the smearing of the bands due to rotational disorder, which provides an additional level of complexity. An accurate assessment of the size of the gap requires taking the rotational disorder into account, as discussed next.

The predicted EDC obtained from the rotational disorder model described above, labelled \Ga\ , with different simulated gaps are shown (blue solid lines) in Fig. \ref{Theory} (a). The most significant change with gap size is reducing the depth of the minima at the Dirac crossing ( $\sim$ -0.3 eV).  The experimental  EDC curve is overlayed (red, dashed line) on three EDC curves (E$_{\rm g}$ = 270, 250 and 230 meV). The best fit between experiment ( fig. \ref{Cu111} (b)) and theory ( fig. \ref{Cu111} (c)) is obtained when a gap of $\sim$ 250 meV is used in the model. This is an appreciably larger gap than was found for graphene on 1 ML Cu/Ni(111)\cite{Varykhalov:2010fd}, E$_{\rm g}$ = 180 meV, which indicates a stronger interaction between the graphene and single crystal Cu. We find that the doping level and bandgap does not vary around the $\pi$ band arc. 

In contrast to the above analysis, a mere visual inspection of the data in Fig. 4b and 5a would suggest a much larger gap, around 350 to 400 meV. This demonstrates that the rotational smearing of the spectral function due to azimuthal disorder is to artificially enhance the gap size by up to 60\% .

The graphene on Cu(100) Fermi surface is shown in Fig. \ref{Cu100} (a), where the graphene ring is twice as broad as for graphene on Cu(111), Fig. \ref{Cu111} (a). The spectral function along the white line in Fig. \ref{Cu100} (a), shown in Fig. \ref{Cu100} (b) indicates a shoulder (green, dash dot dot line) offset from the main band (blue, dash dot line). The TB fit (green, dash dot dot line) describes this shoulder well in \ref{Cu100} (a) and (b). Although the main peak in Fig. \ref{Cu100} can be described by the same TB fit, a higher Dirac crossing offset (e2p = -0.6 eV) is obtained by the fitting. The extra width of the Fermi surface can therefore be ascribed to two distinct doping levels for graphene on single crystal Cu(100). The origin of the two doping levels will be discussed in the next section, however it is important to note that one doping level is the same as the graphene on single crystal Cu(111) case (\Ga\ in Fig. \ref{Cu111} and \ga\ in Fig. \ref{Cu100}) and an extra doping level most likely due to some difference in the sample(\gb\ in Fig. \ref{Cu100}).

Similar to graphene on single crystal Cu(111), Fig. \ref{Cu111} (d) and (e), the EDC and MDC plots in Fig. \ref{Cu100} (d) and (e) indicate intensity in the region of the Dirac crossings. To determine the size of the gap in both doping levels ( \ga\ and \gb\ ) it is once again necessary to compare the EDC curves at the K point from the rotational disorder model and the experimental data. In this case the model is extended by creating the 3D spectral functions, as used in the graphene on Cu(111) case, for each of the two doping levels indicated by the TB bare bands in Fig. \ref{Cu100} and summing them together. The EDC curves at the K point (blue, solid lines) are shown for different simulated gaps in the \gb\ (Fig. \ref{Theory} (b)) and \ga\ (Fig. \ref{Theory} (c)) bands, respectively. In Fig. \ref{Theory} (b) the \ga\ band gap is set to 250 meV and in Fig. \ref{Theory} (c) the \gb\ band gap is set to 350 meV. The overlying experimental EDC (red dashed lines) give the best fit for a gap in the \gb\ band (Fig. \ref{Theory} (b)) of 350 meV and a gap in the \ga\ band (Fig. \ref{Theory} (c)) of 250 meV. 

The model spectral function obtained based on the two TB bare bands from Fig. \ref{Cu100} and the band gaps $E_g$(\ga) = 250 meV and $E_g$(\gb) = 350 meV is shown in Fig. \ref{Cu100} (c), and matches well with the experimental spectral function in Fig. \ref{Cu100} (b). A structural origin for the two doping levels, where two facets on the surface produce a momentum offset of two similarly doped graphene domains, was also considered. The model spectral function for such a system did not reproduce the shape of the experimental EDC plots in Fig. \ref{Theory} (b) and (c), and therefore is discounted. Such facetted features where also not observed in STM measurements of Rasool et al.\cite{Rasool:2011eh}

Summarizing the ARPES data, graphene bands Cu(111) and Cu(100) ( \Ga\ and \ga , resp.) were observed with identical gap ($E_g$=250 meV) and doping level ($\epsilon_{2p}$ = -0.3 eV).  A second graphene band (\gb ) was observed on Cu(100), with larger gap ($\sim 350$ meV) and doping level ($\sim 0.6$ eV). This implies that the interaction of the graphene with the substrate layer is substantially stronger in the \gb\ doping regions.

The interactions between graphene and metal substrates has been modeled extensively using DFT L(S)DA theory\cite{Khomyakov:2009ui} and by considering van der Waals interactions using variations on vdW-DF theory\cite{Hamada:2010je}. Khomyakov et al.\cite{Khomyakov:2009ui}, in addition to calculating the band structure, investigated the effect of varying the bond distance between the graphene layer and a metallic substrate. For graphene on Cu(111) they found that varying the bond distance from 3 \AA\ to 4 \AA\ can shift the Dirac crossing from $\sim$ -0.5 eV to $\sim$ +0.25 eV. The close agreement of the Dirac crossing for graphene on single crystal Cu(111), graphene on single crystal Cu(100) and graphene on 1ML Cu/Ni(111)\cite{Varykhalov:2010fd} therefore indicates that the graphene-Cu bond distance is similar in all cases.

On both Cu(111) and (100) the spectral function does not vary appreciably around the arc of the graphene $\pi$ band. That is, doping level and bandgap do not change with grapheneÕs in-plain orientation. Furthermore, the Dirac crossing values (dopings) and bandgaps are similar on the two Cu surfaces. So grapheneÕs electronic structure is not very sensitive to the physical and electronic structures of the Cu surfaces or the precise alignment of the film/substrate lattices. These observations suggest relatively weak interactions between the film and the Cu. But there are sufficient interactions to dope the graphene and open a bandgap by breaking the symmetry of the graphene lattice. The larger band gap on the Cu single crystals compared to graphene on 1ML Cu/Ni(111)\cite{Varykhalov:2010fd} may result from greater substrate-induced symmetry breaking. This is in contrast to the DFT calculations, which predict only a small band gap (11 meV\cite{Giovannetti:2007jc}.

\subsection{Air exposed films}
To gain insight into the origin of the two doping levels observed for graphene on Cu(100), the sample was re-examined by LEEM/LEED after the ARPES experiments. After degassing at about 250 $^{\rm o}$C, the LEEM images were indistinguishable from the as-grown film. However, Fig. \ref{Cu100LEEM} (c) reveals new LEED features -- four weak, radial lines rotated 45$^{\rm o}$ relative to the directions of the first-order Cu spots. At the ends of the lines are weak spots (see green arrow) separated by $\sqrt{2}$  times the length of the Cu reciprocal lattice. This pattern can be interpreted as induced by the interaction of air with the graphene/Cu(100) system.  Since graphene on other substrates survives air exposure and is cleaned easily with annealing at 300$^{\circ}$ C, it is likely that the effect represents intercalation of an atmospheric species between the graphene and the copper.

As shown by LEED and scanning tunneling microscopy (STM),\cite{Mayer:1986p5395,Jensen:1990p5396} annealing oxygen adsorbed on Cu(100) at about 300$^{\rm o}$C forms two domains with ($\sqrt{2} \times 2\sqrt{2}$)R45$^{\rm o}$ symmetry at a saturation coverage of one half monolayer. Since our samples were annealed to a similar temperature after transferring them through the air to the ARPES chamber, it is likely that we are observing the intercalation of oxygen between the graphene and the Cu(100) surface. In our case, the intercalated oxygen is not well-ordered, giving lines rather than strong LEED spots. 

Auger electron spectroscopy (AES), not shown here, confirmed that the air-exposed sample contained oxygen. Surveying the surface by examining LEED from areas 2 $\mu$m in diameter established that the intercalation was uniform on this length scale. However, we cannot exclude spatial variations in oxygen content at smaller length scales. Thus, we propose that the two doping levels of graphene on Cu(100) observed in ARPES arise from intercalated and non-intercalated regions, which would lead to two different levels of charge transfer to the graphene. The effect of the oxygen on the graphene then also acts to increase the size of the gap from $\sim$ 250 meV to $\sim$ 350 meV. It is interesting that in their study of graphene on 1ML X/Ni(111), with X being Cu, Ag or Au, Varykhalov et al.\cite{Varykhalov:2010fd} also found that the band gap became larger with increasing doping. We see a similar increase in the band gap for the oxygen intercalated region, which has a hgiher doping level.

Even lengthy air exposure did not change the LEEM images and LEED from the graphene /Cu(111) specimen. However, AES revealed the presence of oxygen. Previous studies have established that adsorbed oxygen on Cu(111) does not form ordered structures and changes the work function less than 15 meV.\cite{Jacob:1986p5397} While we cannot totally exclude some oxygen intercalation upon air exposure for graphene/Cu(111), the net effect is a uniform doping of the graphene. 

\subsection{Cu(111) surface state}

\begin{figure}
\includegraphics{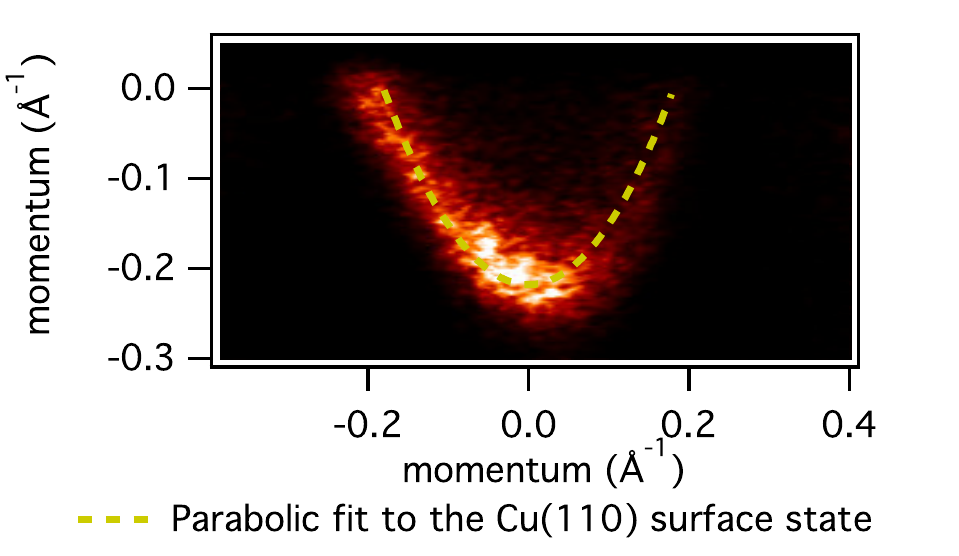}
\caption{\label{Surfacestate}Detailed experimental band structure of the Cu(111) surface state. Overlayed, dashed yellow line, is the parabolic fit used to determine the values in Table \ref{eff_mass}.}
\end{figure}

An important feature of the graphene on Cu(111) Fermi surface (Fig. \ref{Cu111} (a)) is the well known Cu surface state at $\Gamma$ indicated by the small circle (dashed yellow line). Previous measurements\cite{Bendounan:2005p5356} have shown that 1 monolayer of Ag on Cu(111) reduces the bandwidth of this state by a factor of 0.55 and renormalizes the effective mass by a factor of 0.88. The spectral function of the surface state is shown in Fig. \ref{Surfacestate}, with a parabolic fit to the state shown in yellow (small dashed lines) in Fig. \ref{Cu111} (a) and Fig. \ref{Surfacestate}. Parameters from the parabolic fit are used to determine the bandwidth and the Fermi vector presented in Table \ref{eff_mass} along with the values found for pure Cu and 1 ML Ag on Cu by Bendounan et al.\cite{Bendounan:2005p5356}. The bandwidth shift is not accompanied by a Fermi velocity renormalization and is attributed to a simple charge transfer between the graphene and the Cu(111).

As both the surface state and the graphene $\pi $ band have an approximately circular Fermi surface the amount of charge transfer can be related to the change in radius of the Fermi surface in both cases. Such analysis shows that the amount of charge transfer to the surface state is appreciably less than to the graphene $\pi $ bands. It is therefore expected that the majority of the charge transferred to the graphene layer comes from the Bulk Cu and not the surface layer.

The presence of a surface superstructure has been shown to alter the effective mass\cite{Bendounan:2005p5356,Schiller:2005p5402} as well as open a gap in the surface state at the surface brillouin zone boundary ( $\sim$ 0.151 \AA$^{-1}$ for 1 ML Ag on Cu(111)\cite{Bendounan:2005p5356}). The absence of such a gap in the spectral function (Fig. \ref{Surfacestate}) and the similarity of the effective mass from clean Cu(111) , $\sim$ 0.41, and graphene on Cu(111), $\sim$ 0.39, indicates that no surface superstructure exists in this system. This is in agreement with the LEEM data, which indicated an oxygen superstructure on the graphene on Cu(100) sample, but not on the graphene on Cu(111) sample. It is therefore likely the presence of ordered, intercalated oxygen that leads to the second doping level in the graphene on Cu(100) samples, which is not observed for graphene on Cu(111).

\begin{table}
\caption{Cu(111) surface state data. The effective mass is unchanged by the graphene overlayer, in contrast to 1 ML Ag on Cu(111)\cite{Bendounan:2005p5356}.}
\label{eff_mass}\centering%
\begin{tabular*}{8.5cm}{p{3.5cm}p{1.5cm}p{1.5cm}p{1.5cm}}
\hline%
\footnotesize Substrate			& \footnotesize Bandwidth (eV)	& \footnotesize Fermi vector (\AA$^{-1}$)	& \footnotesize Effective mass (m$^\ast/ \rm{m}_{\bar{e}}$) \\\hline%
\footnotesize Cu(111)\cite{Bendounan:2005p5356}			&\footnotesize 0.435	&\footnotesize 0.215	&\footnotesize 0.41						\\
\footnotesize 1ML Ag on Cu(111)\cite{Bendounan:2005p5356}	&\footnotesize 0.241	& \footnotesize 0.151	&\footnotesize 0.36						\\
\footnotesize 1ML G on Cu(111)	& \footnotesize0.218	&\footnotesize 0.150	& \footnotesize0.39								\\ 
\hline
\end{tabular*}
\end{table}


\section{Conclusion}
We investigated graphene grown on single crystal Cu(111) and Cu(100) surfaces. The graphene layers show a high degree of rotational disorder resulting in graphene K points forming a nearly unbroken arc in the diffraction space measurements of both ARPES and LEED measurements, this rotational disorder was also observed in STM measurements\cite{Rasool:2011eh}. The graphene lattice does preferentially align with the Cu(111) lattice, though. The doping level, measured by the offset of the Dirac crossing from the Fermi level ($\sim$ -0.3 eV), was found to be similar for both substrates and to that found for graphene on 1ML Cu/Ni(111) ($\sim$ -0.31 eV). On exposure to air oxygen was found to intercalate under the graphene on the Cu(100) surface forming a superstructure, and consequently a second doping level of the overlaying graphene. The doping level of this second state is higher ($\sim$ -0.6 eV) owing to more charge transfer from the oxygen than the Cu.  The size of the gap induced by interactions with the substrate was found to be larger than for graphene on 1ML Cu/Ni(111) (250 meV and 180 meV respectively) and even larger for the intercalated oxygen regions ($\sim$ 350 meV). Interestingly the interaction between graphene and the Cu(100) and Cu(111) surfaces appears to be similar due to identical doping levels and gaps, with a stronger interaction observed with the intercalated oxygen. Similar to previous studies\cite{Varykhalov:2010fd} we find that the current DFT theory underestimates the band gap ($\sim$11 meV\cite{Giovannetti:2007jc}) by a factor of 30. The Cu surface state at $\Gamma$ on the Cu(111) surface is also found to be protected under the graphene, with only a doping derived binding energy shift being observed.

\begin{acknowledgments}
The Advanced Light Source is supported by the Director, Office of Science, Office of Basic Energy Sciences, of the U.S. Department of Energy under Contract No. DE-AC02-05CH11231. Work at Sandia was supported by the Office of Basic Energy Sciences, Division of Materials Sciences and Engineering of the U.S. DOE under Contract No. DE-AC04-94AL85000. A.W. acknowledges support by the Max Planck Society.
\end{acknowledgments}



%
%

%


\bibliography{references}

\end{document}